\documentclass[10pt]{article}

\usepackage[final]{graphics}
\usepackage{graphicx}          	 
\usepackage{bm}                 
\usepackage{amsmath}             
\usepackage{amssymb}
\usepackage{amsfonts}             
\usepackage{verbatim}           
\usepackage{amsthm}             
\usepackage{lineno}

\usepackage{mathtools}
\usepackage{relsize}

\numberwithin{equation}{section}	
\usepackage{hyperref}

\theoremstyle{plain}             

\newtheorem{theorem}{Theorem}[section]

\newtheorem{corollary}[theorem]{Corollary}

\theoremstyle{definition}

\newtheorem{remark}[theorem]{Remark}

\input epsf

\def\calC{{{\cal C}}}
\def\sign{{\rm sign}}

\def\dsp{\displaystyle}

\def\RR{\mathbb R}
\def\dist{{\rm dist}}
\begin{document}

\title{A faster and more accurate algorithm for calculating population genetics statistics requiring sums of Stirling numbers of the first kind}

\author{
Swaine L. Chen\footnote{
Corresponding author. Infectious Diseases Programme, Department of Medicine, Yong Loo Lin School of Medicine, National University of Singapore, Singapore 119228, Singapore \& Infectious Diseases Group, Genome Institute of Singapore, Singapore 138672, Singapore. Email: slchen@gis.a-star.edu.sg
}
\and
Nico M. Temme\footnote{IAA, 1825 BD 25, Alkmaar, The Netherlands. Former address: Centrum Wiskunde \& Informatica (CWI), Science Park 123, 1098 XG Amsterdam,  The Netherlands. Email: nico.temme@cwi.nl}
}



\maketitle

\begin{abstract}
Ewen's sampling formula is a foundational theoretical result that connects probability and number theory with molecular genetics and molecular evolution; it was the analytical result required for testing the neutral theory of evolution, and has since been directly or indirectly utilized in a number of population genetics statistics.
Ewen's sampling formula, in turn, is deeply connected to Stirling numbers of the first kind.
Here, we explore the cumulative distribution function of these Stirling numbers, which enables a single direct estimate of the sum, using representations in terms of the incomplete beta function.
This estimator enables an improved method for calculating an asymptotic estimate for one useful statistic, Fu's $F_s$.
By reducing the calculation from a sum of terms involving Stirling numbers to a single estimate, we simultaneously improve accuracy and dramatically increase speed.
\end{abstract}




\section{Introduction}
The dominant paradigm in population genetics is based on a comparison of observed data with parameters derived from a theoretical model \cite{pmid28270526,pmid11595044}.
Specifically for DNA sequences, many techniques have been developed to test for extreme relationships between average sequence diversity (number of DNA differences between individuals) and the number alleles (distinct DNA sequences in the population).
In particular, such methods are widely used to predict selective pressures, where certain mutations confer increased or decreased survival to the next generation \cite{pmid11595044}.
Such selective pressures are relevant for understanding and modeling practical problems such as influenza evolution over time \cite{pmid14726583} and during vaccine production \cite{pmid30561532}; adaptations in human populations, which may impact disease risk \cite{pmid25834723,pmid27821149}; and the emergence of new infectious diseases and outbreaks \cite{pmid27601641}.

Many population genetics tests are therefore formulated as unidimensional test statistics, where the pattern of DNA mutations in a sample of individuals is reduced to a single number \cite{pmid11595044,pmid28270526,pmid9335623}.
Such statistics are heavily informed by combinatorial sampling and probability distribution theories, many of which are built upon the foundational Ewens's sampling formula \cite{pmid4667078}, which describes the expected distribution of the number of alleles in a sample of individuals, given the nucleotide diversity.

Ewens's sampling formula not only was a seminal result for population genetics, but also established connections with combinatorial stochastic processes, algebra, and number theory \cite{crane2016}.
For population genetics, in particular, Ewens's sampling formula provided a key analytical result that finally enabled mathematical tests of the neutral theory of evolution \cite{crane2016,pmid11595044}.
It has given rise to several classical population genetics tests for neutrality, including the Ewens-Watterson test, Slatkin's exact test, Strobeck's $S$, and Fu's $F_s$ \cite{pmid9335623,pmid17246396}.
Calculation of subsets of this distribution are useful for testing deviations of observed data from a null model; such subsets often require the calculation of Stirling numbers of the first kind (hereafter referred to simply as Stirling numbers).
In particular, Fu's $F_s$ has recently been shown to be potentially useful for detecting genetic loci under selection during population expansions (such as an infectious outbreak) both in theory and in practice \cite{pmid27601641}.
However, Stirling numbers rapidly grow large and thus explicit calculation can easily overwhelm the standard floating point range of modern computers.

In previous work, an asymptotic estimator for individual Stirling numbers was used to solve the problem of computing Fu's $F_s$ for large datasets, which are now becoming common due to rapid progress in DNA sequencing technology \cite{Chen:2019:ISN}.
Without such improved numerical methods, Fu's $F_s$ calculations for data sets as small as 170 sequences can cause overflow, preventing the use of these statistics for genome-wide screens of selection.
This algorithm based on estimating individual Stirling numbers solved problems of numerical overflow and underflow, maintained good accuracy, and substantially increased speed compared with other existing software packages \cite{Chen:2019:ISN}.
However, there was still a need to estimate multiple Stirling numbers (up to half the total number of sequences).
Here, we explore the potential for further increasing both accuracy and speed in calculating Fu's $F_s$ by using a single estimator for the entire sum, which involves multiple Stirling numbers.

\section{Methods}
\label{sec:methods}
\subsection{General Definitions and Theory}

We take a population of $n$ individuals, each of which carries a particular DNA sequence $D_i$ (referred to as the allele of individual $i$).
We define a metric, $\dist(D_i, D_j)$ to be the number of positions at which sequence $D_i$ differs from $D_j$.
Then, we denote the average pairwise nucleotide difference as $\theta_\pi$ (hereafter referred to simply as $\theta$), defined as:
\begin{equation}
  \theta = \frac{2}{n(n+1)} \sum_{i=1}^{n-1}\sum_{j=i+1}^n \dist(D_i, D_j).
  \label{eq:intro01}
\end{equation}

We also define a set of unique alleles $Du_i \in \{ D_i \}$ which have the property of $(i \ne j) \implies (\dist(Du_i, Du_j) > 0)$.
The ordinality of $\{ Du_i \}$ is denoted $m$, i.e. the number of distinct alleles in the data set.

Building upon on Ewens's sampling formula \cite{pmid9335623,pmid4667078}, it has been shown that the probability that, for given $n$ and $\theta$, at least $m$ alleles would be found, is
\begin{equation}
  S^\prime_{n,m}(\theta)=\frac{1}{(\theta)_n}\sum_{k= m}^n(-1)^{n-k}S_n^{(k)}\theta^k,\quad \theta>0,
  \label{eq:intro02}
\end{equation}
where $(\theta)_n$ is the Pochhammer symbol, defined by
\begin{equation}
  (\theta)_0=1,\quad (\theta)_n=\theta(\theta+1)\cdots(\theta+n-1)=\frac{\Gamma(\theta+n)}{\Gamma(\theta)}.
  \label{eq:intro03}
\end{equation}
$S_n^{(k)}$ is a Stirling number and is defined by:
\begin{equation}
  (\theta)_n=\sum_{k= 0}^n(-1)^{n-k}S_n^{(k)}\theta^k,
  \label{eq:intro04}
\end{equation}

Fu's $F_s$ is then defined as:
\begin{equation}
  F_s=\ln\frac{S^\prime_{n,m}(\theta)}{1-S^\prime_{n,m}(\theta)}.
  \label{eq:intro05}
\end{equation}
Fu's $F_s$ thus measures the probability of finding a more extreme (equal or higher) number of alleles than actually observed.
It requires computing a sum of terms containing Stirling numbers, which rapidly become large and therefore impractical to calculate explicitly even with modern computers \cite{Chen:2019:ISN}.

Because of the relation in \eqref{eq:intro04}, the statistics quantity $S^\prime_{n,m}(\theta)$ satisfies $0\le S^\prime_{n,m}(\theta)\le 1$.
Also, this relation and \eqref{eq:intro03} show that $(-1)^{n-m}S_n^{(m)}$ are non-negative.
We have the special values
\begin{equation}
\begin{aligned}
 & S_n^{(n)}=1\ (n\ge0), \\
 & S_n^{(0)}=0\ (n\ge1), \\
 & S_n^{(1)}=(-1)^{n-1}(n-1)! \ (n\ge1).
  \label{eq:intro06}
\end{aligned}
\end{equation}
There is a recurrence relation
\begin{equation}
  S_{n+1}^{(k)}=S_{n}^{(k-1)}-nS_{n}^{(k)},
  \label{eq:intro07}
\end{equation}
which easily follows from \eqref{eq:intro04}.
For a concise overview of properties, with a summary of the uniform approximations, see \cite[\S11.3]{Gil:2007:NSF}.

We introduce a complementary relation
\begin{equation}
  T^\prime_{n,m}(\theta)=1-S^\prime_{n,m}(\theta)=\frac{1}{(\theta)_n}\sum_{k= 0}^{m-1}(-1)^{n-k}S_n^{(k)}\theta^k,
  \label{eq:intro08}
\end{equation}
leading to an alternate calculation for Fu's $F_s$ of
\begin{equation}
  F_s=\ln\frac{S^\prime_{n,m}(\theta)}{1-S^\prime_{n,m}(\theta)}=\ln\frac{1-T^\prime_{n,m}(\theta)}{T^\prime_{n,m}(\theta)}.
  \label{eq:intro09}
\end{equation}

The recent algorithm considered in \cite{Chen:2019:ISN} is based on asymptotic estimates of $S_n^{(m)}$ derived in \cite{Temme:1993:AES}, which are valid for large values of $n$, with unrestricted values of $m\in (0,n)$.
It avoids the use of the recursion relation given in \eqref{eq:intro07}.

In the present paper we derive an integral representation of 
$S^\prime_{n,m}(\theta)$ and of the complementary function $T^\prime_{n,m}(\theta)$, for which we can use the same asymptotic approach as for the Stirling numbers without calculating the Stirling numbers themselves.
From the integral representation we also obtain a representation in which the incomplete beta function occurs as the main approximant.
In this way we have a convenient representation, which is available as well for many classical cumulative distribution functions.
We show numerical tests based on a first-order asymptotic approximation, which includes the incomplete beta function.
In a future paper we give more details on the complete asymptotic expansion of $S^\prime_{n,m}(\theta)$, and, in addition, we will consider an inversion problem for large $n$ and $m$: to find $\theta$ either from the equation $S^\prime_{n,m}(\theta)=s$, when $s\in(0,1)$ is given, or from the equation $F_s=f$, when $f\in\RR$ is given.

\subsection{Remarks on computing \boldmath{$S^\prime_{n,m}(\theta)$} }

When computing the quantity $F_s$ defined in \eqref{eq:intro05}, numerical instability may happen when $S^\prime_{n,m}(\theta)$ is close to 1.
In that case, the computation of $1-S^\prime$ suffers from cancellation of digits.
For example, take $n= 100$, $\theta= 39.37$, $m= 31$.
Then $S^\prime_{n,m}(\theta)\doteq0.99872$, and $F_s$ becomes about $6.6561$ when using the first relation in \eqref{eq:intro09}.
However, when we calculate $T^\prime_{n,m}(\theta)= 0.002689$ and use the second relation, then we obtain the more reliable result $F_s\doteq5.9160$.

We conclude that, when $S^\prime_{n,m}(\theta)\ge0.5$, it is better to switch and obtain $T^\prime_{n,m}(\theta)$ from the sum in \eqref{eq:intro08} and $F_s$ using the second relation in \eqref{eq:intro09}.
A simple criterion to decide about this can be based on using the saddle point $z_0$ (see Remark~\ref{rem:rem03} below).

A second point is numerical overflow when $n$ is large, because $S_n^{(m)}$ rapidly becomes large when $m$ is small with respect to $n$.
For example, when $n=10$, $m=5$ we have
\begin{equation}
  S_{10}^{(5)}=-\frac{n!(m+5)(m+4)(3m^2+23m+38)}{11520(m-1)!}=-269325.
  \label{eq:rem01}
\end{equation}
Therefore, it is convenient to scale the Stirling number in the form $S_n^{(k)}/n!$.
In addition, the Pochhammer term $(\theta)_n$ in front of the sum in \eqref{eq:intro02} will also be large with $n$; we have $(1)_n=n!$.

We can write the sum in \eqref{eq:intro02} in the form
\begin{equation}
  S^\prime_{n,m}(\theta)=\frac{n!}{(\theta)_n}\sum_{k= m}^n(-1)^{n-k}\widehat{S}_n^{(k)}\theta^k,\quad \widehat{S}_n^{(k)}=\frac{{S}_n^{(k)}}{n!}.
  \label{eq:rem02}
\end{equation}
Leading to a corresponding modification in the recurrence relation in \eqref{eq:intro07} for the scaled Stirling numbers:
\begin{equation}
  \widehat{S}_{n+1}^{(m)}=\frac{1}{n+1}\left(\widehat{S}_{n}^{(m-1)}-n\widehat{S}_{n}^{(m)}\right).
  \label{eq:rem03}
\end{equation}

To control overflow, we can consider the ratio
\begin{equation}
  f_n(\theta)=\frac{n!}{(\theta)_n}=\frac{\Gamma(n+1)\,\Gamma(\theta)}{\Gamma(\theta+n)}.
  \label{eq:rem04}
\end{equation}
This function satisfies $f_n(\theta)\le1$ if $\theta\ge1$.
For small values of $n$ we can use recursion in the form
\begin{equation}
  f_{n+1}(\theta)=\frac{n+1}{n+\theta}f_n(\theta),\quad n=0,1,2,\ldots, \quad f_0(\theta)=1.
  \label{eq:rem05}
\end{equation}
For large values of $n$ and all $\theta>0$, we can use a representation based on asymptotic forms of the gamma function.

It should be observed that using the recursions in \eqref{eq:intro07} and \eqref{eq:rem03} is a rather tedious process when $n$ is large.
For example, when we use it to obtain $S_{100}^{(m)}$ for all $m\in(0,100]$, we need all previous $S_n^{(m)}$ with $n\le99$ for all $m\in(0,n]$.
A table look-up for $\widehat{S}_{n+1}^{(m)}$ in floating point form may be a solution.
When $n$ is large enough, the algorithm mentioned in \cite{Chen:2019:ISN} evaluates each needed Stirling number by using the asymptotic approximation derived in \cite{Temme:1993:AES}.

\subsection{Data Availability}

Code implementing the new estimator for Fu's $F_s$ in R is available at https://github.com/swainechen/hfufs.

\section{Results and Discussion}

\subsection{Analytical Results}

The new algorithm is based on the following results, which we describe in two theorems.
\begin{theorem}\label{thm:theo01}
The statistics quantity $S_{n+1,m+1}^\prime(\theta)$ introduced in \eqref{eq:intro02} has the representation as an integral in the complex $z$-plane
\begin{equation}
  S_{n+1,m+1}^\prime(\theta)=\frac{\theta^{m}}{(\theta+1)_n}\frac{1}{2\pi i}\int_{\calC_R} \frac{(z+1)_n}{z^{m}}\,\frac{dz}{z-\theta},\quad R > \theta,
  \label{eq:math01}
\end{equation}
where $n$ and $m$ are positive integers, $0\le m\le n$, $\theta$ is a real positive number, and $\calC_R$ is a circle at the origin with radius $R>\theta$.
The symbol $(\alpha)_n$ denotes the Pochhammer symbol introduced in \eqref{eq:intro03}.
 \end{theorem}
Observe that we have raised in $S_{n,m}^\prime(\theta)$ the parameters $n$ and $m$ with unity; this is convenient in the mathematical analysis. The proof of this theorem will be given in the Appendix (\nameref{sec:app01}).

\begin{corollary}\label{cor:cor01}
The complementary quantity $T_{n+1,m+1}^\prime(\theta)$ introduced in \eqref{eq:intro08} has the representation
\begin{equation}
  T_{n+1,m+1}^\prime(\theta)=\frac{\theta^{m}}{(\theta+1)_n}\frac{1}{2\pi i}\int_{\calC_R} \frac{(z+1)_n}{z^{m}}\,\frac{dz}{\theta-z},\quad R < \theta.
  \label{eq:math02}
\end{equation}
\end{corollary}

The main asymptotic result is given in the second theorem.
\begin{theorem}\label{thm:theo02}
$S_{n+1,m+1}^\prime(\theta)$ has the representation 
\begin{equation}
\begin{aligned}
 & S_{n+1,m+1}^\prime(\theta)=I_x(m, n-m+1)+R_{n+1,m+1}^\prime(\theta) \\
 & x= {\frac{\tau}{1+\tau}},\quad \tau >0,
  \label{eq:math03}
\end{aligned}
\end{equation}
where $I_x(p,q)$ is the incomplete beta function defined by
\begin{equation}\label{eq:math04}
I_x(p,q)=\frac{1}{B(p,q)}\int_0^x t^{p-1}(1-t)^{q-1}\,dt,
\end{equation}
with
\begin{equation}\label{eq:math05}
0 < x < 1,\quad p >0, \quad q> 0, \quad B(p,q)=\frac{\Gamma(p)\Gamma(q)}{\Gamma(p+q)}.
\end{equation}
The term $R_{n+1,m+1}^\prime(\theta)$ is a function of which we give a one-term approximation in \eqref{eq:math16}.
\end{theorem}

\begin{corollary}\label{cor:cor02}
The complementary quantity $T_{n+1,m+1}^\prime(\theta)$ has the representation
\begin{equation}
\begin{aligned}
 & T_{n+1,m+1}^\prime(\theta)=I_{1-x}(n-m+1,m)-R_{n+1,m+1}^\prime(\theta), \\
 & 1-x= {\frac{1}{1+\tau}}.
  \label{eq:math06}
\end{aligned}
\end{equation}
\end{corollary}

This follows from Theorem~\ref{thm:theo01} and the complementary relation of the incomplete beta function
\begin{equation}\label{eq:math07}
I_x(p,q)=1-I_{1-x}(q,p).
\end{equation}

Note also that the incomplete beta function in \eqref{eq:math03} has the representation (see \cite[\S8.17(i)]{Paris:2010:INC})
\begin{equation}\label{eq:math19}
I_\frac{\tau}{1+\tau}(m,n-m+1)=(1+\tau)^{-n}\sum_{j=m}^n\binom{n}{j}\tau^j,
\end{equation}
and from the complementary relation in \eqref{eq:math07} it follows that the function in \eqref{eq:math06} has the expansion
\begin{equation}\label{eq:math20}
I_\frac{1}{1+\tau}(n-m+1,m)=(1+\tau)^{-n}\sum_{j=0}^{m-1}\binom{n}{j}\tau^j.
\end{equation}

The representation in this theorem in terms of the probability function $I_x(p,q)$ shows the characteristic role of $S_{n,m}^\prime(\theta)$ as a cumulative distribution function of the Stirling numbers. The representation can also be viewed as an asymptotic representation in which the incomplete beta function is the main approximant.

The proof of Theorem~\ref{thm:theo02} can be found in the Appendix (\nameref{sec:app02}), but we give here some preliminary information about functions used in the proof to explain the definition of the parameter $\tau$ in \eqref{eq:math03}.
It is a function of $\theta$ and arises in certain transformations of the integral given in Theorem~\ref{thm:theo01}.
For this we need the function
\begin{equation}
\begin{aligned}
  \phi(z) & = \ln\left( (z+1)_n\right)-m\ln z \\
          & = \ln\Gamma(z+n+1)-\ln\Gamma(z+1)-m\ln z,
  \label{eq:math08}
\end{aligned}
\end{equation}
and its derivative
\begin{equation}
  \phi^\prime(z)=\psi(z+n+1)-\psi(z+1)-\frac{m}{ z}=0, \quad \psi(z)=\frac{\Gamma^\prime(z)}{\Gamma(z)}.
  \label{eq:math09}
\end{equation}
With the function $\phi(z)$ we can write \eqref{eq:math01} in the form
\begin{equation}
  S_{n+1,m+1}^\prime(\theta)=\frac{e^{-\phi(\theta)}}{2\pi i}\int_{\calC_R} e^{\phi(z)}\frac{dz}{z-\theta},\quad R > \theta.
  \label{eq:math10}
\end{equation}
Then the saddle point of the integral in \eqref{eq:math10} follows from the equation
\begin{equation}
  \phi^\prime(z)=\psi(z+n+1)-\psi(z+1)-\frac{m}{ z}=0, \quad \psi(z)=\frac{\Gamma^\prime(z)}{\Gamma(z)}.
  \label{eq:math11}
\end{equation}
There is a positive saddle point $z_0$ when $0 < m <n$.

\begin{figure}[htbp]
\centering
\includegraphics[width=\linewidth]{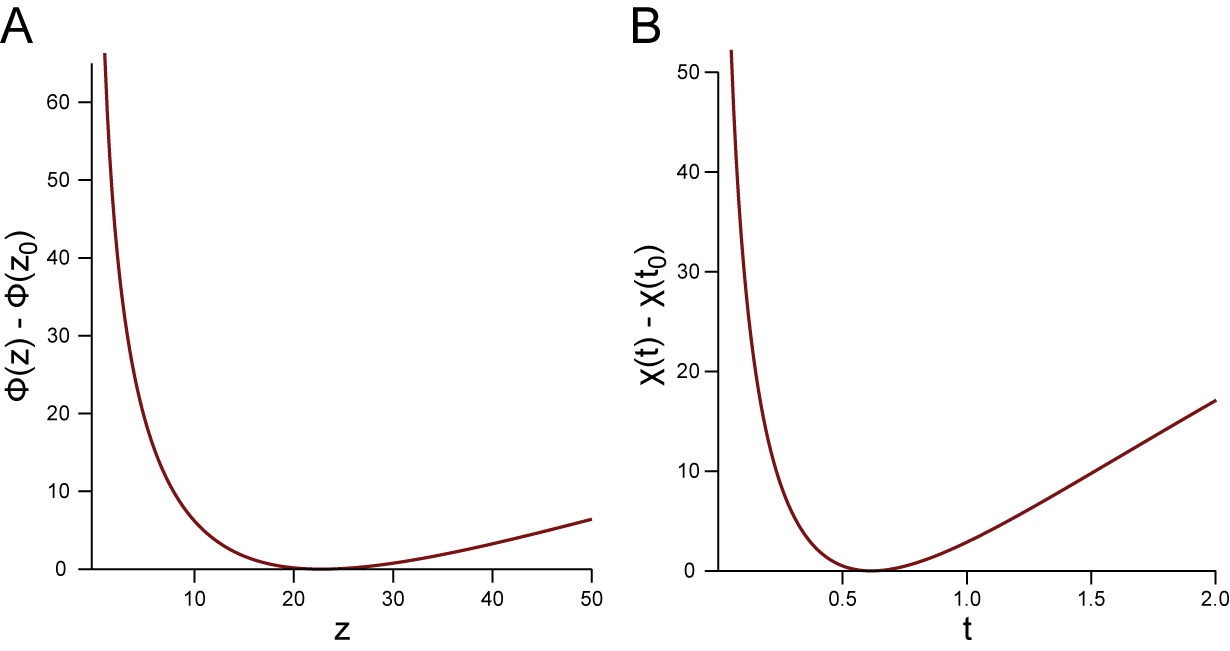} 
\caption{Graphs of $\phi(z)-\phi(z_0)$ (A) and $\chi(t)-\chi(t_0)$ (B) for $n=100$, $m=38$, with $z_0\doteq22.81$ and $t_0=\frac{19}{31}\doteq0.61$.}
\label{fig:params}
\end{figure}

Next to these functions we introduce a function for complex values of a variable $t$:
\begin{equation}
\begin{aligned}
 & \chi(t)=n\ln(1+t)-m\ln t, \\
 & \chi^\prime(t)=\frac{(n-m)t-m}{t(1+t)}=(n-m)\frac{t-t_0}{t(1+t)},
  \label{eq:math12}
\end{aligned}
\end{equation}
where $t_0=\frac{m}{n-m}$. These functions are related by
\begin{equation}
  \phi(z)-\phi(z_0)=\chi(t)-\chi(t_0),
  \label{eq:math13}
\end{equation}
with condition $\sign(z-z_0)=\sign(t-t_0)$. In this way, using this relation as a transformation of the variable $z$ to $t$, we can write \eqref{eq:math10} as
\begin{equation}
\begin{aligned}
 & S_{n+1,m+1}^\prime(\theta)=\frac{e^{-\chi(\tau}}{2\pi i}\int_{\calC_R} e^{\chi(t)}f(t)\,dt, \\
 & f(t)=\frac{1}{z-\theta}\frac{dz}{dt}=\frac{1}{z-\theta}\frac{\chi^\prime(t)}{\phi^\prime(z)}.
  \label{eq:math14}
\end{aligned}
\end{equation}

The parameter $\tau$ in Theorem~\ref{thm:theo02} is defined as the positive solution of the equation
\begin{equation}
  \phi(\theta)-\phi(z_0)=\chi(\tau)-\chi(t_0),\quad \sign(\theta-z_0)=\sign(\tau-t_0).
  \label{eq:math15}
\end{equation}

In Figure~\ref{fig:params} we show the graphs of $\phi(z)-\phi(z_0)$ (Figure~\ref{fig:params}A) and $\chi(t)-\chi(t_0)$ (Figure~\ref{fig:params}B) for $n=100$, $m=38$.
For these values the saddle points are $z_0\doteq22.81$ and $t_0=\frac{19}{31}\doteq0.61$.
The sign condition $\sign(z-z_0)=\sign(t-t_0)$ for the relation in \eqref{eq:math13} means the left branches of the convex curves correspond with functions values for $z\in(0,z_0]$ and $t\in(0,t_0]$, and the right branches with values for $z\in[z_0,\infty)$ and $t\in[t_0,\infty)$.
Clearly, we have a one-to-one relation between the positive $z$ and $t$-variables.

A first-order approximation of the function $R_{n+1,m+1}^\prime(\theta)$ in \eqref{eq:math03} and \eqref{eq:math06} reads
\begin{equation}
\begin{aligned}
 & R_{n+1,m+1}^\prime(\theta)\sim e^{-\chi(\tau)}\binom{n}{m-1}g(t_0), \\
 & n\to\infty, \quad 0 < m < n,
  \label{eq:math16}
\end{aligned}
\end{equation}
where
\begin{equation}
  g(t_0)= f(t_0)-\frac{1}{t_0-\tau}, \quad f(t_0)=\frac{1}{z_0-\theta}\sqrt{\frac{\chi^{\prime\prime}(t_0)}{\phi^{\prime\prime}(z_0)}},
  \label{eq:math17}
\end{equation}
and the function $f(t)$ is defined in \eqref{eq:math14}.
The value $f(t_0)$ follows from evaluating $dz/dt$ (see \eqref{eq:math14}) at $t_0$, by observing that both functions $\phi^\prime(z)$ and $\chi^\prime(t)$ vanish when $t\to t_0$ (hence, $z\to z_0$).
Then, l'H{\^o}pital's rule can be used to obtain $f(t_0)$.

\begin{figure}[htbp]
\centering
\includegraphics[width=\linewidth]{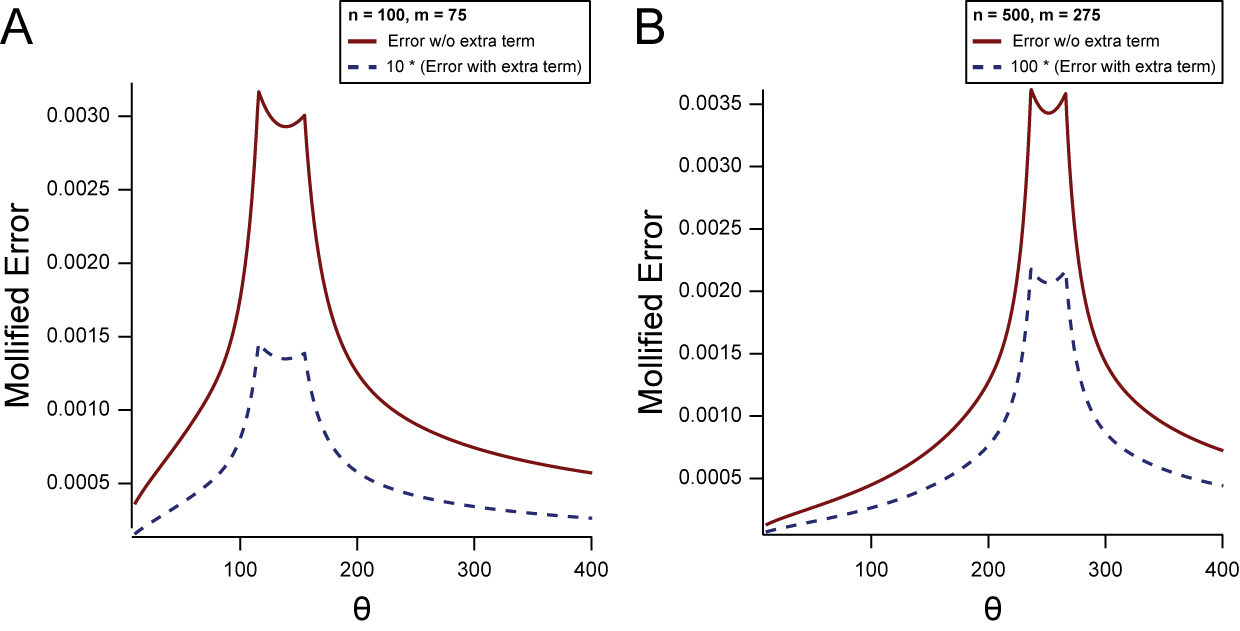}
\caption{Mollified error in estimating Fu's $F_s$ for $\theta\in(10,400)$, $m=75$ and $n=100$ (A) and for $m=275$ and $n=500$ (B). The data for the dashed curves are multiplied by a factor of 10 (A) and 100 (B), to make the error curves visible in the figures. Refer to the text for further details.}
\label{fig:errorcurves}
\end{figure}

In Figure~\ref{fig:errorcurves} we show the error curves $\delta(F_s,\widetilde F_s)$ in \eqref{eq:math18} for Fu's $F_s$ \eqref{eq:intro09} for $\theta\in[10,400]$.
We show examples for $n=100$, $m=75$ (Figure~\ref{fig:errorcurves}A) and $n=500$, $m=275$ (Figure~\ref{fig:errorcurves}B).
The solid curves are for $F_s$ when using $S_{n+1,m+1}^\prime(\theta)\sim I_{\tau/(1+\tau)}(m,n-m+1)$, the dashed curves when using $S_{n+1,m+1}^\prime(\theta)\sim I_{\tau/(1+\tau)}(m,n-m+1)+R_{n+1,m+1}^\prime(\theta)$ with the asymptotic estimate given in \eqref{eq:math16}.
For ease of visualization, the error $\delta(F_s,\widetilde F_s)$ has been multiplied by a factor 10 or 100 in Figure~\ref{fig:errorcurves}.
We have used the following {\em mollified error} function
\begin{equation}\label{eq:math18}
\delta(F_s,\widetilde F_s)=\left\vert\frac{F_s-\widetilde F_s}{{\rm max}(\vert F_s\vert,1)}\right\vert,
\end{equation}
where $\widetilde F_s$ is the approximation of $F_s$.
This mollified error is exactly the relative error unless $\vert F_s\vert$ is small.
Because $F_s$ will vanish when $S_{n+1,m+1}^\prime(\theta)=\frac12$ (which also means that $\theta$ is near the transition value $z_0\doteq 137.98$ (in Figure~\ref{fig:errorcurves}A) and $z_0\doteq251.58$ (in Figure~\ref{fig:errorcurves}B) (see Remark~\ref{rem:rem03})), we cannot use relative error for all $\theta>0$.
This explains the non-smooth curves in Figure~\ref{fig:errorcurves}.

The final estimator is based on the representations in \eqref{eq:math03} and \eqref{eq:math06} and the first order approximation in \eqref{eq:math16}, which are used to calculate Fu's $F_s$ with one of the two relations in \eqref{eq:intro09} depending on whether $S^\prime_{n,m}(\theta)\ge0.5$, decided as described above.

\subsection{Implementation and Numerical Results}

We first summarize the steps to compute Fu's $F_s$ by using \eqref{eq:intro09} and the first-order approximations (see \eqref{eq:math16} and \eqref{eq:math03} or \eqref{eq:math06})
\begin{equation}\label{eq:num01}
S_{n+1,m+1}^\prime(\theta)\sim I_{\frac{\tau}{1+\tau}}(m, n-m+1) +e^{-\chi(\tau)}\binom{n}{m-1}\ g(t_0),
\end{equation}
or
\begin{equation}\label{eq:num02}
T_{n+1,m+1}^\prime(\theta)\sim I_{\frac{1}{1+\tau}}(n-m+1,m)- e^{-\chi(\tau)}\binom{n}{m-1}\ g(t_0),
 \end{equation}
for large $n$, $m\in(0,n)$ and $\theta>0$.

\begin{enumerate}
\item
Compute the saddle point $z_0$, the positive zero of $\phi^\prime(z)$; see \eqref{eq:math11}.
\item
With $t_0=m/(n-m)$, the positive zero of $\chi^\prime(t)$ (see \eqref{eq:math12}), compute $\tau$, the solution of the equation (see \eqref{eq:math15})
\begin{equation}\label{eq:num03}
\chi(\tau)=\chi(t_0)+\phi(\theta)-\phi(z_0),
\end{equation}
with $\phi(z)$ defined in \eqref{eq:math08} and $\chi(t)$ defined in \eqref{eq:math12}. When $\theta=z_0$ there is one solution $\tau= t_0$. When $\tau \ne t_0$ there are two positive solutions, and we take the one that satisfies the condition $\sign(\theta-z_0)=\sign(\tau-t_0)$.
\item 
When $\theta < z_0$, hence $\tau < t_0$, compute the approximation of $S_{n+1,m+1}^\prime(\theta)$ by using \eqref{eq:num01}, and $F_s$ from the first relation in \eqref{eq:intro09}.
\item 
When $\theta > z_0$, hence, $\tau > t_0$, compute the approximation of $T_{n+1,m+1}^\prime(\theta)$ by using \eqref{eq:num02}, and $F_s$ from the second relation in \eqref{eq:intro09}.
\end{enumerate}

\renewcommand{\arraystretch}{1.2}
\begin{table}
\caption{
Relative errors in the computation of $F_s$ defined in \eqref{eq:intro05}
 using the asymptotic estimator in \eqref{eq:num01}.
\label{tab:tab01}}
$$
\begin{array}{rcrrc}
n/m\quad & \theta & {F_s, \rm asymptotic}& {F_s, \rm exact\quad}&{\rm rel.
error}  \\
\hline
25/20        \quad &  9.39 &       -6.83168 &     -6.8294578 & 0.33\times 10^{-3}\\
50/31        \quad &  9.61 &     -10.13052 &   -10.1290263 & 0.15\times 10^{-3}\\
100/40      \quad &  9.37 &     -10.23064 &   -10.2298131 & 0.81\times 10^{-4}\\
250/67      \quad &  8.96 &     -26.41607 &   -26.4155959 & 0.18\times 10^{-4}\\
500/95      \quad &  9.04 &     -46.76268 & -46.76238956 & 0.63\times 10^{-5}\\
1000/152  \quad &  9.07 &   -112.42500 & -112.4248080 & 0.17\times 10^{-5}\\
2001/213  \quad &  9.03 &   -192.21835 & -192.2182390 & 0.60\times 10^{-6}\\
\hline
\end{array}
$$
\end{table}
\renewcommand{\arraystretch}{1.0}

In Table~\ref{tab:tab01} we show the relative errors in the computation of $F_s$ defined in \eqref{eq:intro05}.
The values of $n$, $m$, and $\theta$ correspond with those in Table~1 of \cite{Chen:2019:ISN}.
The asymptotic result is from \eqref{eq:num01}.
Computations were done with Maple, with Digits = 16.
The ``exact" values were obtained by using Maple's code for $Stirling1(n,m)$, which computes the Stirling numbers of the first kind.

\begin{figure}[htbp]
\centering
\includegraphics[width=\linewidth]{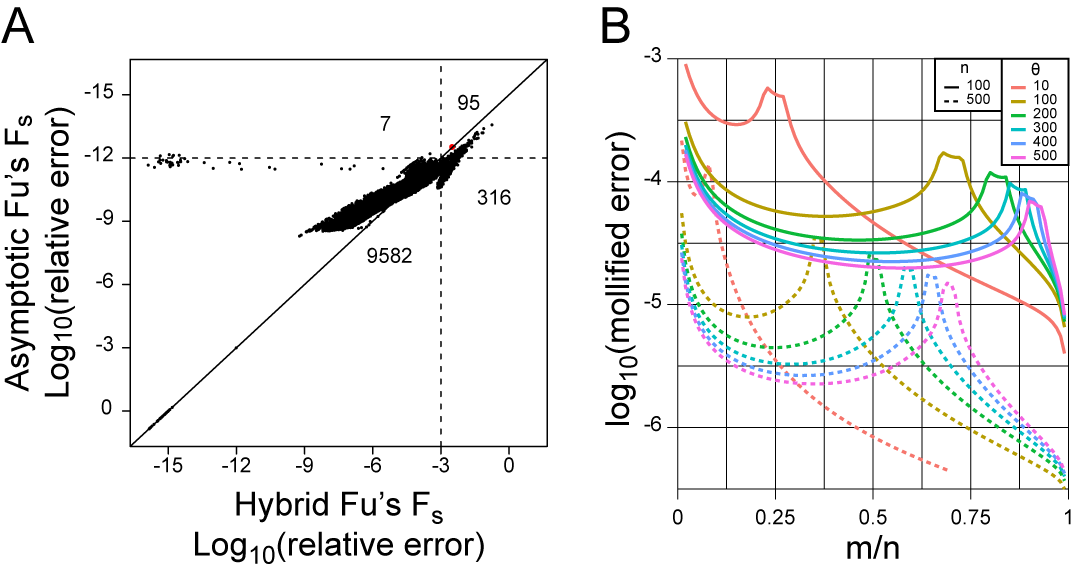}
\caption{(A) Comparison of relative error of the estimator from \cite{Chen:2019:ISN} and the single term asymptotic estimator in \eqref{eq:num01}. Relative error for each is calculated against the arbitrary precision implementation described in \cite{Chen:2019:ISN}. In total, 10,000 calculations were performed with $n$ randomly sampled from a uniform distribution between 50 and 500; $m$ between 2 and $n$; and $\theta$ between 1 and 50. A solid diagonal line is drawn at $y = x$. Dotted lines are drawn at a relative error of 0.001. Numbers within each quadrant defined by the dotted lines indicate the number of points in each quadrant. The red dot indicates the one case where the relative error was $>0.001$ and the error of \eqref{eq:num01} was greater than the estimator from \cite{Chen:2019:ISN}. (B) Comparison of mollified error (\eqref{eq:math18}) as a function of $m$. For this plot, we fixed $n=100$ (solid lines) or 500 (dotted lines) and $\theta\in(10,500)$ (as indicated by different line colors).}
\label{fig:accuracy}
\end{figure}

\begin{figure}[htbp]
\centering
\includegraphics[width=\linewidth]{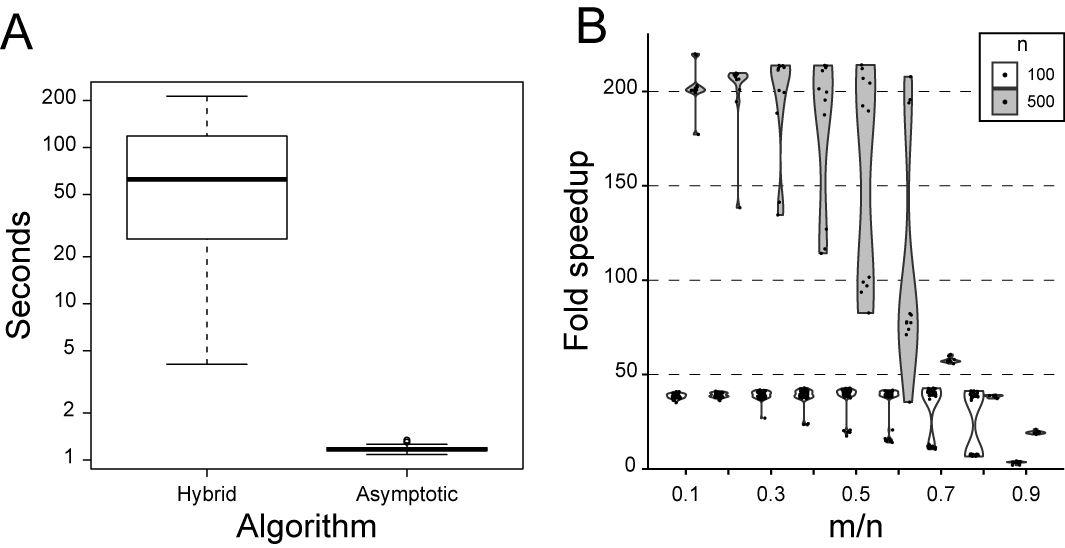}
\caption{(A) Comparison of run times between the hybrid algorithm from \cite{Chen:2019:ISN} and the single term asymptotic estimator in \eqref{eq:num01}. 100 iterations were run, each with 10,000 calculations; the time elapsed for each set of 10,000 calculations was recorded and plotted here. The same set of parameters were used for each algorithm. The order of running the algorithms was alternated with each iteration. The dark horizontal line indicates the median, the box indicates the first and third quartiles, the whiskers are drawn at 1.5x the interquartile range, and outliers are represented by open circles. The median for the hybrid algorithm is 62.64 s; the median for the asymptotic algorithm is 1.17 s. (B) Detailed benchmarking for $n=100$ (open violins) or 500 (gray violins), $m\in(0.1n, 0.2n, ..., 0.9n)$, and $\theta\in(10,500)$. Fold speedup (ratio of the time taken for the hybrid calculator to that taken for the aysmptotic estimator) is plotted on the y-axis. Each dot represents one set of parameters; the violin plots summarize the density of points on the y-axis. Times were calculated for 100 iterations of each estimator for the same parameter values.}
\label{fig:timing}
\end{figure}

We additionally performed a comparison with the recently published algorithm in \cite{Chen:2019:ISN}.
We performed 10,000 calculations with each algorithm and compared the results with an exact calculator.
As expected, since the previous algorithm required estimating a Stirling number for each term of the sum, while the current asymptotic estimate directly calculates the sum, both error and compute speed were improved.
Relative error for the single term estimate in \eqref{eq:num01} was well controlled at $<0.001$ for nearly 99\% of the calculations; for 411 calculations where the previous hybrid estimator had an error $>0.001$, the estimate in \eqref{eq:num01} was more accurate in all but one case ($n=157, m=4, \theta=43.59732$; 3.08e-3 relative accuracy using \cite{Chen:2019:ISN}; 3.32e-3 relative accuracy using \eqref{eq:num01}) (Figure~\ref{fig:accuracy}).
Further analysis of the relative error demonstrated that it peaks at intermediate values of $m/n$, depending on $\theta$.
These correspond to parameter choices near the transition values $m = m_0$, where $t$ approaches $t_0$ and $z$ approaches $z_0$ in the calculation; notably, they remain well controlled (all values $< 0.001$ mollified error) regardless of $\theta$.
The asymptotic behavior (lower relative error) can also be seen as both $n$ and $m$ increase in the right panel of Figure~\ref{fig:accuracy}.

The fewer calculations led to a clear improvement in calculation speed (median 54.6x faster; Figure~\ref{fig:timing}).
The speedup also depends on the parameter choices; in general, the speed advantage is greater when the hybrid calculator requires many calculations (namely, when $m$ is small relative to $n$, as the hybrid calculator performs the sum in \eqref{eq:intro02}) (Figure~\ref{fig:timing}).

\section{Conclusion}

The rapid growth of sequencing data has been an enormous boon to population genetics and the study of evolution.
Traditional population genetics statistics are still in common use today.
The statistics Fu's $F_s$ and Strobeck's $S$ have been difficult to calculate on modern, large data sets using previous methods; we now further improve both accuracy and speed for the calculation of Fu's $F_s$ such data sets, using the main estimator in \eqref{eq:num01}.
Our plan for a paper about the ability to invert the calculation provides additional future directions in understanding the performance of these statistics.
Therefore, the methods used herein may be useful for the development of new statistics that more effectively capture different types of selection.
\section*{Acknowledgments}
The authors tank the reviewers for their helpful comments and suggestions.\\
SLC acknowledges Shyam Prabhakar and members of the Chen lab for fruitful discussions. NMT acknowledges CWI, Amsterdam, for scientific support; he thanks Edgardo Cheb--Terrab (Maplesoft) for his help with the Maple code. \\
SLC was supported by the National Medical Research Council, Ministry of Health, Singapore (grant numbers NMRC/OFIRG/0009/2016 and\\ NMRC/CIRG/1467/2017).\\
NMT was supported by the Ministerio de Ciencia e Innovaci\'on, Spain, projects MTM2015-67142-P (MINECO/FEDER, UE) and \\
PGC2018-098279-B-I00 (MCIU/AEI/FEDER, UE).
The authors affirm that all data necessary for confirming the conclusions of the article are present within the article, figures, and tables.


\section{Appendix}\label{sec:app}

\subsection{Proof of Theorem~\ref{thm:theo01}}
\label{sec:app01}
We use the integral representation of the Stirling numbers that follows from the definition given in \eqref{eq:intro04}.
That is, by using Cauchy's formula,
\begin{equation}
  (-1)^{n-m}S_{n}^{(m)}=\frac{1}{2\pi i}\int_{\calC_R}(z)_n \frac{dz}{z^{m+1}},
  \label{eq:app01}
\end{equation}
where ${\calC_R}$ is a circle around the origin with radius $R$.
We can take $R$ as large as we like.
As in \cite[\S3]{Temme:1993:AES}, it is convenient to proceed with
\begin{equation}
  (-1)^{n-m}S_{n+1}^{(m+1)}=\frac{1}{2\pi i}\int_{\calC_R}(z+1)_n \frac{dz}{z^{m+1}}.
  \label{eq:app02}
\end{equation}
Using the definition of $S_{n,m}^\prime(\theta)$ in \eqref{eq:intro02} we have
\begin{equation}
  \begin{aligned}
  S_{n+1,m+1}^\prime(\theta) & = \frac{1}{(\theta)_{n+1}}\sum_{k= m+1}^{n+1} (-1)^{n+1-k}S_{n+1}^{(k)}\theta^k \\
  & = \frac{1}{(\theta+1)_{n}}\sum_{k= m}^{n} (-1)^{n-k}S_{n+1}^{(k+1)}\theta^k.
  \end{aligned}
  \label{eq:app03}
\end{equation}
and using \eqref{eq:app02} we obtain
\begin{equation}
  S_{n+1,m+1}^\prime(\theta)=\frac{1}{(\theta+1)_{n}}\sum_{k= m}^{n} \frac{\theta^k}{2\pi i}\int_{\calC_R} \frac{(z+1)_n}{z^{k+1}}\,dz.
  \label{eq:app04}
\end{equation}

We can take $R>\theta$ to have $\vert \theta/z\vert<1$ on the circle ${\calC_R}$, and we can perform the summation to $\infty$, because all terms with $k > n$ do not give contributions.
In this way we obtain the requested integral representation
\begin{equation}
  S_{n+1,m+1}^\prime(\theta)=\frac{\theta^{m}}{(\theta+1)_n}\frac{1}{2\pi i}\int_{\calC_R} \frac{(z+1)_n}{z^{m}}\,\frac{dz}{z-\theta},\quad R > \theta.
  \label{eqapp05}
\end{equation}
This concludes the proof of Theorem~\ref{thm:theo01}.

Corollary~\ref{cor:cor01} now follows by using the theory of integrals of analytical functions on complex contours.
We have assumed that $R>\theta$, but we can take $R < \theta$ while picking up the residue at $z=\theta$. The result is
\begin{equation}
  S_{n+1,m+1}^\prime(\theta)=1-\frac{\theta^{m}}{(\theta+1)_n}\frac{1}{2\pi i}\int_{\calC_R} \frac{(z+1)_n}{z^{m}}\,\frac{dz}{\theta-z}, \quad R < \theta.
  \label{eq:app06}
\end{equation}
This gives the relation in Corollary~\ref{cor:cor01}.

\begin{remark}\label{rem:rem03}
When $\theta$ crosses the value $z_0$, $S_{n+1,m+1}^\prime(\theta)$ becomes (almost) $\frac12$.
Especially when the parameters $m$ and $n$ are large, $S_{n+1,m+1}^\prime(\theta)$ starts with very small values for small $\theta$, becomes close to $\frac12$ when $\theta=z_0$, and quickly becomes 1 as $\theta$ increases.
We call $z_0$ the {\em transition value} for $\theta$.

For fixed values of $n$ and $\theta$, there is also a transition value for $m$; refer to this transition value as $m_0$.
When $n$ is large, $S_{n+1,m+1}^\prime(\theta)$ starts at values near 1 for small $m$, it becomes about $\frac12$ when $m$ nears $m_0$, and it becomes quickly small as $m\to n$.
\end{remark}

\subsection{Proof of Theorem~\ref{thm:theo02}}
\label{sec:app02}
The relation in \eqref{eq:math13} between the functions $\phi(z)$ (see \eqref{eq:math08}) and $\chi(t)$ (see \eqref{eq:math12}) can be used as a transformation of the variable $z$ to $t$, as in \cite[\S3]{Temme:1993:AES}.
The result is the integral representation in \eqref{eq:math14}.
In Figure~\ref{fig:params} we have shown the relationship between $z$ and $t$.

The function $f(t)$ in \eqref{eq:math14} has a pole in the $t$-domain; refer to this pole as $t=\tau$.
This then corresponds with the pole at $z=\theta$ in the $z$-domain.
The relation between $\tau$ and $\theta$ follows from the transformation given in \eqref{eq:math13}.
In other words, $\tau$ is defined by the equation 
\begin{equation}
  \phi(\theta)-\phi(z_0)=\chi(\tau)-\chi(t_0), \quad \sign(\theta-z_0)=\sign(\tau-t_0),
  \label{eq:app07}
\end{equation}
where the sign-convention follows from the one used for \eqref{eq:math13}.
We can express the existence of the pole of the function $f(t)$ defined in \eqref{eq:math14} by writing
\begin{equation}
  f(t)=\frac{1}{z-\theta}\frac{dz}{dt}=\frac{t-\tau}{z-\theta}\frac{dz}{dt}\frac{1}{t-\tau}.
  \label{eq:app08}
\end{equation}

In asymptotic analysis, the presence of such a pole is of great interest, especial when (in the $t$-domain) the saddle point (here $t_0$) is close to a pole (here $\tau$), or even when these points coalesce.
See, for example, \cite[Chapter~21]{Temme:2015:AMI}.
Usually, the error function is introduced to handle the asymptotic analysis; in the present case, we use an incomplete beta function.
We split off the pole from $f(t)$ and write
\begin{equation}
  f(t)=\frac{A}{t-\tau}+g(t),
  \label{eq:app09}
\end{equation}
where we assume that $g(t)$ is well defined at $t=\tau$.
To find $A$ we use the analytical relation in \eqref{eq:math13} between $t$ and $z$, in particular at $z=\theta$ (or $t=\tau$).
Applying l'H{\^o}pital's rule, we find that $\dsp{\frac{t-\tau}{z-\theta}\frac{dz}{dt}\to1}$ as $t\to\tau$, which gives $A=1$.
Hence, substituting this form of $f(t)$ in \eqref{eq:math14}, we find
\begin{equation}
  \begin{aligned}
  S_{n+1,m+1}^\prime(\theta) & = \frac{e^{-\chi(\tau)}}{2\pi i}\int_{\calC_S} \frac{(t+1)^n}{t^{m}}\,\frac{dt}{t-\tau}
  & + \frac{e^{-\chi(\tau)}}{2\pi i}\int_{\calC} \frac{(t+1)^n}{t^{m}}g(t)\,dt.
  \end{aligned}
  \label{eq:app10}
\end{equation}
The radius of the circle $\calC_S$ in the first integral is larger than $\tau$.
For the second integral, we take a circle $\calC$ around the origin such that the singularities of $g(t)$ are outside the circle.

In \nameref{sec:app03} below, we prove that the first integral in \eqref{eq:app10} can be evaluated in terms of the incomplete beta function as shown in Theorem~\ref{thm:theo02}.
We can then write \eqref{eq:app10} as
\begin{equation}
  S_{n+1,m+1}^\prime(\theta)=I_{\frac{\tau}{1+\tau}}(m, n-m+1)+R_{n+1,m+1}^\prime(\theta),
  \label{eq:app11}
\end{equation}
where
\begin{equation}
  R_{n+1,m+1}^\prime(\theta) = \frac{e^{-\chi(\tau)}}{2\pi i}\int_{{\cal C}_S} \frac{(t+1)^n}{t^{m}}g(t)\,dt.
  \label{eq:app12}
\end{equation}
A first-order approximation of \eqref{eq:app12} follows from replacing $g(t)$ by its value at the saddle point $t_0$.
This gives
\begin{equation}
  R_{n+1,m+1}^\prime(\theta)\sim e^{-\chi(\tau)}\binom{n}{m-1}\ g(t_0),
  \label{eq:app13}
\end{equation}
where
\begin{equation}
  g(t_0) = f(t_0)-\frac{1}{t_0-\tau},\quad f(t_0)=\frac{1}{z_0-\theta}\sqrt{\frac{\chi^{\prime\prime}(t_0)}{\phi^{\prime\prime}(z_0)}}.
  \label{eq:app14}
\end{equation}
This expression of $f(t_0)$ follows from \eqref{eq:math17}.
In a future publication we will give details about the complete asymptotic expansion of the term $R_{n+1,m+1}^\prime(\theta)$.

\subsection{Proof of the incomplete beta relation}
\label{sec:app03}
We give a proof of the claim that the incomplete beta function in \eqref{eq:app11} equals the first integral in \eqref{eq:app10}.
That is,
\begin{equation}
  \frac{e^{-\chi(\tau)}}{2\pi i}\int_{\calC_S} \frac{(t+1)^n}{t^{m}}\,\frac{dt}{t-\tau}=I_{\frac{\tau}{1+\tau}}(m, n-m+1),
  \label{eq:app15}
\end{equation}
where ${\calC_S} $ is a circle at the origin with radius larger than $\tau$.
We have, using the definition of $\chi(t)$ in \eqref{eq:math12},
\begin{equation}
  \begin{aligned}
  \frac{e^{-\chi(\tau)}}{2\pi i}\int_{\calC_S} \frac{(t+1)^n}{t^{m}}\,\frac{dt}{t-\tau} & = \frac{(1+\tau)^{-n}\tau^m}{2\pi i}\int_{\calC_S} \frac{(t+1)^n}{t^{m+1}}\,\frac{dt}{1-\tau/t} \\
        & = (1+\tau)^{-n}\sum_{k=0}^{n-m}\tau^{k+m} \frac{1}{2\pi i}\int_{\calC_S} \frac{(t+1)^n}{t^{m+k+1}}\,dt \\
        & =(1+\tau)^{-n}\sum_{k=0}^{n-m} \tau^{k+m} \binom{n}{m+k} \\
        & =(1+\tau)^{-n}\sum_{j=m}^n \tau^{j} \binom{n}{j},
  \end{aligned}
  \label{eq:app16}
\end{equation}
which is the relation in \eqref{eq:math19}.
In the second line we have used a finite number of terms of the infinite expansion of $1/(1-\tau/t)$ because terms with $k>n-m$ do not give a contribution.

\end{document}